\documentclass[aps,pra,reprint,superscriptaddress,floatfix]{revtex4-2}

\usepackage[colorlinks=true,urlcolor=blue]{hyperref}

\usepackage{amsmath,amssymb,bm,bigints}
\usepackage{braket}

\usepackage{graphicx}
\graphicspath{{./images/}}

\begin{document}

\title{Topological Phase Diagram of Optimally Shaken Honeycomb Lattices: 
A Dual Perspective from Stroboscopic and Non-Stroboscopic Floquet Hamiltonians}

\author{{\'A}lvaro R. Puente-Uriona}
\affiliation{Centro de F{\'i}sica de Materiales, Universidad del Pa{\'i}s Vasco UPV/EHU, 20018 San Sebasti{\'a}n, Spain}

\author{Giulio Pettini}
\affiliation{Dipartimento di Fisica e Astronomia, Universit\`a di Firenze, and INFN, 50019 Sesto Fiorentino, Italy}

\author{Michele Modugno}
\affiliation{Department of Physics, University of the Basque Country UPV/EHU, 48080 Bilbao, Spain}
\affiliation{IKERBASQUE, Basque Foundation for Science, 48009 Bilbao, Spain}
\affiliation{EHU Quantum Center, University of the Basque Country UPV/EHU, 48940 Leioa, Biscay, Spain}

\date{\today}

\begin{abstract}
We present a direct comparison between the stroboscopic and non-stroboscopic effective approaches for ultracold atoms in shaken honeycomb lattices, focusing specifically on the optimal driving introduced by A. Verdeny and F. Mintert [Phys. Rev. A \textbf{92}, 063615 (2015)]. 
In the fast-driving regime, we compare the effective non-stroboscopic Hamiltonian derived through a perturbative expansion with a non-perturbative calculation of the stroboscopic Floquet Hamiltonian, obtained through a simple non-perturbative numerical approach. 
 We show that while some of the tunneling parameters are inherently model-dependent, the topological properties of the system remains robust, as expected. 
 Using the same numerical approach we compute the topological phase diagram, arguing that it is most effectively represented in terms of the physical parameters characterizing the driving and the bare Hamiltonian -- 
 parameters directly accessible in experiments -- rather than the emergent tunneling parameters, 
 that depend on the model representation.  
\end{abstract}
\maketitle

\section{Introduction}

Recent years have witnessed the flourishing of the so-called \textit{Floquet engineering} in the context of of ultracold quantum gases, namely the ability to experimentally manipulate and control the behavior of these systems by applying time-periodic driving forces \cite{eckardt2005,kierig2008,eckardt2010,kitagawa2010,koghee2012,sticlet2012,struck2012,hauke2012,salger2013,verdeny2013,delplace2013,jotzu2014,goldman2014,baur2014,holthaus2015,bukov2015,eckardt2017,bukov2015a,verdeny2015a,verdeny2015,eckardt2015,creffield2016,plekhanov2017,modugno2017,messer2018,pieplow2018,pieplow2019,cooper2019,boulier2019,harper2020,wintersperger2020,sun2020,dicarli2023,mateos2023,mateos2023b,schnell2023,lin2023,arrouas2023}. This approach takes advantage of the Floquet theory, which describes the behavior of quantum systems under the influence of time-periodic perturbations (for a review see, e.g., Refs. \cite{holthaus2015,bukov2015,eckardt2017}). The key idea behind all this is to engineer the periodic driving force in such a way that it effectively modifies the behavior of the system on timescales much larger than the modulation period. By controlling the parameters of the driving force, one can tailor the energy spectrum of the system, induce artificial gauge fields, create topological states of matter, and realize novel quantum phases that are not possible in static systems. Nowadays, many research laboratories worldwide have the capability to experimentally implement such engineering techniques \cite{kierig2008,salger2013,jotzu2014,messer2018,boulier2019,wintersperger2020,arrouas2023}. 

From a theoretical perspective, different approaches are available to describe such periodically driven systems. A natural choice is provided by the original Floquet theory, in which the evolution operator is factorized in the product of a time-periodic evolution operator and of a one-cycle evolution operator, see, e.g., Ref. \cite {holthaus2015}.  Through the latter one can introduce an \textit{effective} Floquet Hamiltonian $\hat{H}_{F}[t_{0}]$ \cite{jotzu2014,modugno2017}, providing  an effective description of the longtime dynamics of the system. It corresponds to the full evolution of the system on a whole period $T$, from an initial time $t_{0}$ to $t_{0}+T$.
For this reason, it is also called \textit{stroboscopic} Floquet Hamiltonian \cite{bukov2015}. Though $\hat{H}_{F}[t_{0}]$ depends on the choice of the time $t_{0}$ which defines the beginning of the stroboscopic driving period,  it is otherwise time-independent, in the sense that it does not describe the micromotion within each period $T$. 
It is noteworthy that the energy spectrum and topological properties of the effective Hamiltonian do not depend on the choice of the initial time $t_{0}$, since two effective Hamiltonians for different initial times are related through a unitary gauge transformation, 
$\hat{H}_{F}[t_{0}']=\hat{U}(t_{0}', t_{0})\hat{H}_{F}[t_{0}]\hat{U}^{-1}(t_{0}',t_{0})$. The choice of the initial time $t_{0}$ only affects the local properties of the Hamiltonian, i.e., the actual values of the effective tunneling coefficients, but not the general structure of the Hamiltonian. In principle, since the choice of $t_{0}$ is a gauge choice, all the Hamiltonians $\hat{H}_{F}[t_{0}]$ must be gauge equivalent to some fixed Floquet Hamiltonian $\hat{H}_{F}$, that is independent of $t_{0}$ \cite{bukov2015}.

In general, an exact expression for the effective Hamiltonian $\hat{H}_{F}[t_0]$ cannot be found. However, from its formal definition, $\hat{H}_{F}[t_{0}]\equiv(i/T)\ln[\hat{U}(t_{0}+T,t_{0})]$, it can be calculated either perturbatively or numerically. The perturbative approach amounts to the so-called Magnus expansion (see, e.g., Refs. \cite{bukov2015,blanes2009}), that is an analytic Taylor expansion of $\hat{H}_{F}[t_{0}]$ in powers of $\omega^{-1}$ (with $\omega\equiv2\pi/T$).
On the numerical side, a convenient approach -- that will be employed in this paper -- consists in factorizing the time-ordered product over infinitesimal time intervals and handling the resulting expression in the basis of the Pauli matrices \cite{jotzu2014,modugno2017}. 

A different but related strategy is the one introduced by Goldman and Dalibard in Ref. \cite{goldman2014}  (see also \cite{bukov2015} and references therein), in which the evolution operator is factorized in three terms, one corresponding to the effects associated with the initial phase of the modulation, another one to the longtime dynamics of the system -- to which one can associate an effective Hamiltonian $\hat{H}_{\textrm{eff}}$ --, and finally one accounting for the micromotion. 
Notably, $\hat{H}_{\textrm{eff}}$ is independent of $t_{0}$ by definition. Therefore, with respect to the standard Floquet approach described above, it offers the advantage that all quantities are manifestly gauge invariant, by definition. However, in general it cannot be defined by an explicit analytical expression but only in terms of a perturbative expansion in powers of the inverse driving frequency $\omega^{-1}$ \cite{goldman2014}. The latter is the analogue of the Magnus expansion for $\hat{H}_{F}[t_0]$. Indeed,  one can establish a formal correspondence between $\hat{H}_{F}[t_0]$ and $\hat{H}_{\textrm{eff}}$, at each order of the perturbative expansion in $\omega^{-1}$, see, e.g., Ref. \cite{bukov2015}. The other side of the coin is that without an explicit analytical expression defining the Hamiltonian $\hat{H}_{\textrm{eff}}$, it is generally not possible to compute it numerically in a non-perturbative manner.

To gain a deeper understanding of these concepts, in this paper we provide a direct comparison between the two approaches, by considering a specific illustrative example. To this end, we chose the proposal by Verdeny and Mintert for realizing a tunable Chern insulator \cite{verdeny2015}, consisting in a system of noninteracting ultracold atoms loaded in a honeycomb lattice and subjected to a bichromatic \textit{optimal} driving \footnote{For another insightful comparison between the stroboscopic and non-stroboscopic approaches, in the context of the Harper-Hofstadter model, see Ref. \cite{bukov2015a}.}. As in the original proposal, the model is considered here within the tight-binding approximation. This system presents interesting similarities with the Haldane model \cite{haldane1988}, a paradigmatic lattice model of a topological insulator, whose phase diagram has been recently  explored in experiments with ultracold fermionic atoms in periodically driven optical lattices \cite{jotzu2014} (see also Ref. \cite{modugno2017}). 

Specifically, here we focus the analysis on the fast-driving regime, in which the first-order expression of $\hat{H}_{\textrm{eff}}$ derived in Ref. \cite{verdeny2015} is expected to be fully justified. We compute the effective (stroboscopic) Floquet Hamiltonian $\hat{H}_{F}[t_0]$ by means of the non-perturbative numerical approach of Ref. \cite{modugno2017}, that allows for a straightforward calculation of the tunneling coefficients and of the topological phase diagram. 
These results are then compared to the first-order prediction for $\hat{H}_{\textrm{eff}}$ \cite{verdeny2015}. In particular, we show that within the parameter regime considered in this paper, the perturbative approach well reproduces the topological properties of the model. We also discuss how to conveniently draw the phase diagram in terms 
of the model parameters associated with the driving and the explicit breaking of the inversion symmetry (IS), without relying on the specific form of the first-order effective Hamiltonian. 
This is done while considering the fact that IS is dynamically broken by this class of optimal driving. Our discussion also encompasses other interesting aspects, such as the gauge and model dependence of the tunneling coefficients, which, however, is not relevant to the topological properties.

The paper is organized as follows. In Sec. \ref{sec:models}, we outline the theoretical framework for driven lattice models, specifically focusing on the honeycomb lattice with nearest-neighbor hoppings under the influence of bichromatic sinusoidal driving. In Sec. \ref{sec:floquet}, we discuss the stroboscopic and non-stroboscopic approaches for computing the effective Floquet Hamiltonian. First, in Sec. \ref{sec:perturbative}, we revisit the perturbative formulation for constructing the `optimal' effective Hamiltonian as presented in Ref. \cite{verdeny2015}. Then, in Sec. \ref{sec:numerical}, we present the methodology for constructing the stroboscopic Floquet Hamiltonian using a non-perturbative numerical approach. We also provide results for the onsite energies and tunneling coefficients of the model, along with a discussion of analogies and differences between the two approaches. Section \ref{sec:topology} is dedicated to the discussion of the topological phase diagram. Finally, we summarize our findings and present concluding remarks in Sec. \ref{sec:conclusions}.

\section{Driven lattice models}
\label{sec:models}

Let us consider a general undriven lattice model described by the \textit{bare} Hamiltonian 
\begin{equation}
\label{eq:und_hamil}
\hat{H}_{b} = \sum_{\nu,\nu'}\sum_{\bm{j},\bm{j}'}t^{\bm{j}-\bm{j}'}_{\nu\nu'}{\hat{c}}^{\dagger}_{\bm{j}\nu}{\hat{c}}_{\bm{j}'\nu'},
\end{equation}
where the indices $\bm{j},\;\bm{j'}$ run over the unit cells of the Bravais lattice while the indices $\nu,\;\nu'$ run over the different sites of a given cell. The coefficients $t^{\bm{j}-\bm{j}'}_{\nu\nu'}$ represent the hopping matrix elements in the model, namely the tunneling coefficients.

If the lattice is subjected to a rigid time-periodic driving $\bm{r}_{lat}(t)$, a particle of mass $m$ loaded in the lattice feel an inertial force $\bm{F}(t)=-m\ddot{\bm{r}}_{lat} \equiv\dot{\bm{q}}(t)$, with $\bm{F}(t+T)=\bm{F}(t)$. Therefore, the total Hamiltonian of the system may be expressed as \cite{jotzu2014,modugno2017}
\begin{equation}
    \hat{H}(t) = \hat{H}_{b} + \sum_{\nu,\nu'}\sum_{\bm{j},\bm{j}'}\bm{F}(t)\cdot\bm{r}_{\bm{j}\nu} \hat{c}^{\dagger}_{\bm{j}\nu}{\hat{c}}_{\bm{j}'\nu'},
\end{equation}
that, in the co-moving frame, can be conveniently written in the compact form as
\begin{align}
    \hat{H}_{lat}(t) &= \sum_{\nu,\nu'}\sum_{\bm{j},\bm{j}'}t^{\bm{j}-\bm{j}'}_{\nu\nu'}
    e^{i\bm{q}(t)\cdot(\bm{r}_{\bm{j}\nu}-\bm{r}_{\bm{j}'\nu'})/\hbar}
    {\hat{c}}^{\dagger}_{\bm{j}\nu}{\hat{c}}_{\bm{j}'\nu'}
    \nonumber\\
    &\equiv \sum_{\nu,\nu'}\sum_{\bm{j},\bm{j}'}t^{\bm{j}-\bm{j}'}_{\nu\nu'}(t)
    {\hat{c}}^{\dagger}_{\bm{j}\nu}{\hat{c}}_{\bm{j}'\nu'},
    \label{eq:tdep}
\end{align}
where
\begin{equation}
   t^{\bm{j}-\bm{j}'}_{\nu\nu'}(t)\equiv
   t^{\bm{j}-\bm{j}'}_{\nu\nu'}
    e^{i\bm{q}(t)\cdot(\bm{r}_{\bm{j}\nu}-\bm{r}_{\bm{j}'\nu'})/\hbar}
    \label{eq:tdep_tun}
\end{equation}
and
\begin{equation}
\bm{q}(t) = \int_{t_{b}}^{t}\bm{F}(t') dt' -\frac{1}{T}\int_{0}^{T}\!\!\!\!dt\int_{t_{b}}^{t}\bm{F}(t') dt',
\label{eq:qt2}
\end{equation}
under the condition of vanishing net transferred momentum (over a period), with $t_{b}$ being an arbitrary integration bound.
In the following we shall work in this representation (namely, the co-moving frame), omitting the index $\textit{lat}$ for easiness of notations. We can also fix $t_{b}=0$, without loss of generality.

When one is interested in the dynamics of the system over a time scale much larger than the driving period $T$, an effective time-independent Hamiltonian may be introduced in terms of the one-period time-evolution operator, the so called effective Floquet Hamiltonian (see, e.g., Ref. \cite{modugno2017} and references therein),
\begin{equation}
\label{eq:eff_f_h}
\begin{split}
     \hat{H}_F[t_{0}] = & \frac{i\hbar}{T}\text{log}\left[\hat{U}(t_{0},t_{0}+T)\right] \\ = & \frac{i\hbar}{T}\text{log}
     \left[\mathcal{T}e^{-\frac{i}{\hbar}\int_{t_{0}}^{t_{0}+T}\hat{H}(t')dt'}\right],
\end{split}
\end{equation}
where $\mathcal{T}$ represents the time-ordering operator. 
The above expression should be interpreted in the context of its Taylor expansion. 
Specifically, by introducing the dimensionless time parameter $\tau\equiv\omega t$, 
the evolution operator can be elegantly expressed as follows,
\begin{align}
&{\cal T}e^{-\displaystyle\frac{i}{\hbar}
\int_{\tau_{0}}^{\tau_{0}+2\pi}\!\!\!\!\!\!\!\!\hat{H}(\tau)d\tau}\equiv 
1+ 
\\
\nonumber
&\quad\sum_{n=1}^{+\infty} \frac{(-i)^{n}}{n!}\frac{1}{(\hbar\omega)^{n}}
\int\! d\tau_{1}\cdots \int\! d\tau_{n} {\cal T}\left[\hat{H}(\tau_{1})\cdots \hat{H}(\tau_{n})\right],
\end{align}
representing a series expansion in $1/\omega$, commonly known in the literature as the Magnus expansion \cite{magnus1954, bukov2015}.
Notice that the operator $\hat{H}_F[t_{0}]$ is invariant under the transformation $\hat{U}(t_{0},t_{0}+T) \longrightarrow \hat{U}(t_{0},t_{0}+T)e^{i 2m\pi }$ (with $m\in \mathbb{Z}$), so that its eigenvalues, the so called \textit{quasienergies}, are defined modulo an integer multiple of $\hbar\omega$ \cite{holthaus2015}.
We also recall that the energy spectrum and topological properties do not depend on the initial time $t_{0}$, which only affects the local properties of the Hamiltonian. 
Owing to this, we can set $t_{0}=0$, without loss of generality. The dependence on $t_{0}$ of the tunneling coefficients within the stroboscopic approach will be briefly discussed later on, in Sec. \ref{sec:numerical}.

In the following, we shall focus on the model considered in Ref. \cite{verdeny2015}, that we review below.

\subsection{Optimally shaken honeycomb lattices}
\label{sec:optimal}

\begin{figure}
	\centerline{\includegraphics[width=0.8\columnwidth]{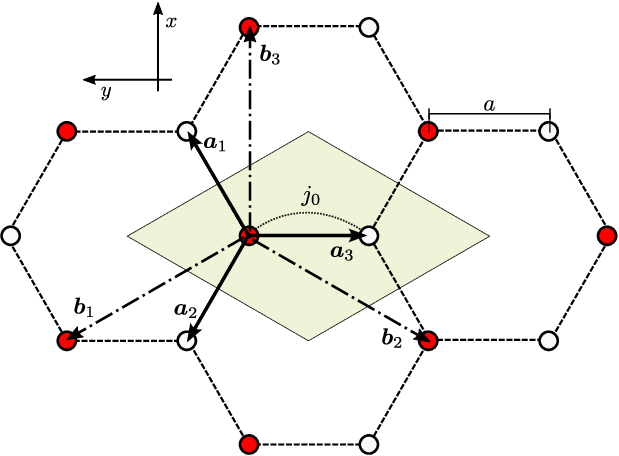}}
	\caption{Sketch of the honeycomb lattice. The unit cell is highlighted in light green, and the sites A and B are denoted by the filled (red) and empty (white) circles, respectively. 
 NN sites are connected by the vectors $\bm{a}_1$, $\bm{a}_2$ and $\bm{a}_3$, with $j_0$ being the corresponding tunneling rate.
 The vectors $\bm{b}_1$, $\bm{b}_2$ and $\bm{b}_3$ connect the different unit cells in the Bravais lattice. Notice that there exists a specific relation between the two sets of vectors, e.g., $\bm{b}_1=\bm{a}_2-\bm{a}_3$; the other relations are obtained by a cyclic permutation of the indices.}
	\label{fig:honeycomb}
\end{figure}

\textit{Bare Hamiltonian.} The undriven system, represented by the Hamiltonian $\hat{H}_{b}$ [see Eq. \eqref{eq:und_hamil}], consists of a honeycomb lattice with two inequivalent sites per unit cell as shown in Fig. \ref{fig:honeycomb}. The geometry of the lattice is identified by two sets of vectors, $\bm{a}_i$ and $\bm{b}_i$ ($i=1,2,3$), that connect nearest-neighbour (NN) sites (of type $A$ and $B$) and neighbouring unit cells, respectively. The model is characterized by an onsite energy offset $2\delta$ (where the onsite energies are $E_{A,B}=\pm\delta$) and an isotropic NN hopping matrix element between sites A and B, represented by the tunneling coefficient $j_0$. All higher order tunneling coefficients are assumed to be vanishing (they are supposed to be negligible, from the physical point of view).

In reciprocal space (see Appendix \ref{appendix1}), the system Hamiltonian is represented by a $2\times2$ matrix $h_{\nu\nu'}(\bm{k})$ that can be conveniently written in a compact form, by using the basis formed by the $2\times 2$ identity matrix, $I$, and of the three Pauli matrices, $\sigma_{i}$ \cite{modugno2017,verdeny2015}. Namely, 
\begin{equation}
 h(\bm{k})=h_{0}(\bm{k})I+\bm{h}(\bm{k})\cdot\bm{\sigma},
 \label{eq:paulidecomp}
\end{equation}
with ${\bm{h}}\equiv(h_{1},h_{2},h_{3})$. 
In the present case, owing to the vanishing of the tunneling coefficients beyond NN and to the fact that $E_A+E_B=0$, we have $h_0(\bm{k})\equiv0$. Remarkably, we shall see that this property -- that constitutes a distinctive difference with respect to the original Haldane model -- is preserved in the presence of the driving, so that $h_0(\bm{k})$ can be omitted from the following discussion. Therefore, the spectrum of $h(\bm{k})$ can be readily written as
\begin{equation}
\epsilon_{\pm}=\pm|\bm{h}|.
\label{eq:spectrum}
\end{equation}
In reciprocal space, the bare Hamiltonian has the following structure 
\begin{align}
h_{1}(\bm{k}) &= j_{0}{\rm{Re}}[Z(\bm{k})],
\\
h_{2}(\bm{k}) &= -j_{0}{\rm{Im}}[Z(\bm{k})],
\\
h_{3}(\bm{k}) &= \delta,
\label{eq:h3undriven}
\end{align}
where the function
\begin{equation}
    Z(\bm{k})=1+e^{i\bm{k}\cdot\bm{b}_{1}}+e^{-i\bm{k}\cdot\bm{b}_{2}}
\end{equation}
is fixed by the lattice geometry. 

\begin{figure}
	\centering
	\includegraphics[width=0.6\columnwidth]{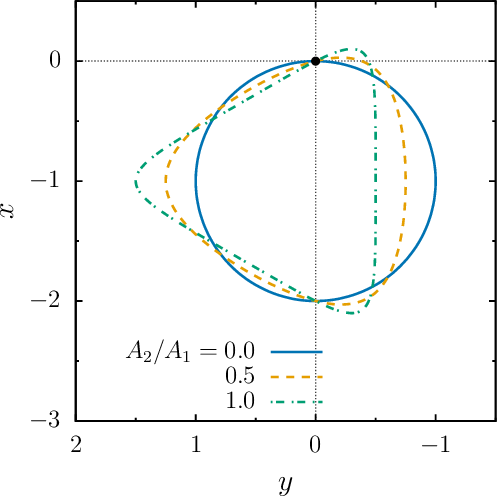}
	\caption{Trajectory of the lattice ($\ddot{\bm{r}}_{lat} =-\bm{F}(t)/m$) under the influence of the force \eqref{opt_forc}, for some values of the ratio $A_2/A_1$. 
    The path is given over a whole period; the thick dot at $(x,y)=(0,0)$ indicates the initial position, at $t=0$. Lengths ($x,y$) are given in units of $A_1/(m\omega^2)$. 
    The orientation of the axis is the same as in Fig. \ref{fig:honeycomb}.
    Notice that when $A_2\neq0$, IS (with respect to the trajectory center) is always explicitly broken.
 However, the trajectory profiles always exhibit at least $C_{3v}$ symmetry. }
	\label{fig:driving}
\end{figure}

\textit{Lattice driving.} To induce topologically nontrivial phases, a time-reversal symmetry (TRS) breaking driving must be applied. Here, we specifically consider the bichromatic driving proposed in Ref. \cite{verdeny2015}, which is designed to achieve isotropic effective nearest-neighbor (NN) and next-nearest-neighbor (NNN) tunneling rates. Namely,
\begin{equation}
\label{opt_forc}
\begin{split}
\bm{F}(t) = &\left[A_1\cos(\omega t) + A_2\cos(2\omega t - \pi/2)\right] \bm{e}_x + \\
&  \left[A_1\cos(\omega t + \pi/2) + A_2\cos(2\omega t - \pi)\right] \bm{e}_y,
\end{split}
\end{equation}
where $\bm{e}_{x,y}$ are the unit vectors of the $x$ and $y$ axis and $\omega=2\pi/T$ is the fundamental frequency of the driving.
Typical trajectories of the lattice center-of-mass, produced by such a force, are shown in Fig. \ref{fig:driving}.
Note that this driving scheme differs from the monochromatic driving employed by Jotzu \textit{et al.} \cite{jotzu2014} in their experimental realization of the Haldane topological phase diagram. In that case, the breaking of TRS was directly controlled by the phase $\varphi$ of the driving (see also Ref. \cite{modugno2017}).

As anticipated in the introduction, it is important to remark that this class of driving explicitly breaks IS.
This is the reason why the model can exhibit a Haldane-like topological phase diagram, even if the undriven lattice Hamiltonian 
is inversion-symmetric (see Ref. \cite{verdeny2015}). Bearing this in mind, in the following we will refer to the inversion-symmetric ($\delta=0$) and IS-breaking ($\delta\neq0$) cases in reference to the properties of the undriven model, specifically whether the bare onsite energies of sites $A$ and $B$ are degenerate or not.

From Eq. \eqref{eq:tdep_tun} it becomes apparent that the driving leaves the onsite energies unaffected, $t_{AA}^{\bm{0}}(t)=\delta=-t_{BB}^{\bm{0}}(t)$, whereas the NN hopping elements $t_{\nu\nu'}^{\bm{j}-\bm{j}'}(t)$ acquire a time dependence. 
In particular, by defining \cite{verdeny2015}
\begin{equation}
\label{verdeny_tun}
    g_{\bm{a}_i}(t) \equiv j_{0}e^{i\chi_{\bm{a}_i}(t)},
\end{equation}
with
\begin{equation}
    \chi_{\bm{a}_i}(t)\equiv \bm{q}(t)\cdot\bm{a}_i/\hbar,
\end{equation}
the three NN tunneling coefficients corresponding to the hopping from the site $A$ in the central unit cell to three neighbouring sites $B$
(see Fig. \ref{fig:honeycomb}) can be expressed as
\begin{equation}
     t_{BA}^{\bm{0}} = g_{\bm{a}_3}(t),\quad  
    t_{BA}^{\bm{b_1}} =g_{\bm{a}_2}(t),\quad
    t_{BA}^{-\bm{b_2}} = g_{\bm{a}_1}(t).
    \label{eq:tunnelings}
\end{equation}
By combining the above results, the components of the \textit{driven} Hamiltonian in reciprocal space read 
\begin{align}
h_{1}(\bm{k},t) &= j_{0}{\rm{Re}}[e^{-i\chi_{\bm{a}_3}(t)}Z(\bm{k}-\bm{q}(t))],
\\
h_{2}(\bm{k},t) &= -j_{0}{\rm{Im}}[e^{-i\chi_{\bm{a}_3}(t)}Z(\bm{k}-\bm{q}(t))],
\\
h_{3}(\bm{k},t) &= \delta,
\end{align}
with $h_{0}(\bm{k},t)\equiv0$.

\section{Effective Floquet Hamiltonian}
\label{sec:floquet}
In general, an exact expression for the effective Hamiltonian, $\hat{H}_{F}[t_0]$ in  Eq. \eqref{eq:eff_f_h} or $\hat{H}_{\textrm{eff}}$  \cite{goldman2014,bukov2015}, is unavailable. 
However, it can be calculated either perturbatively or numerically, as previously mentioned. 
Different analytical approaches are available in the literature, as discussed in Ref. \cite{bukov2015}.
However, perturbative expansions are often limited to the first order and this may not be sufficient for an accurate quantitative description
in a generic driving regime. In this respect, a non-perturbative numerical approach can be very useful. In the following, we will directly compare these two viewpoints within the framework of the driven lattice model presented in the previous section.

\subsection{Perturbative approach}
\label{sec:perturbative}

We begin by examining the perturbative approach, which amounts to an expansion of $\hat{H}_{\textrm{eff}}$ in powers of $\omega^{-1}$. In particular, we adopt the approach of Verdeny and Mintert \cite{verdeny2015}, which serves as a formal definition of the effective Hamiltonian,
\begin{equation}
    \hat{H}_{\textrm{eff}} \simeq \hat{H}_{\textrm{eff}}^{(0)} + \hat{H}_{\textrm{eff}}^{(1)} = \hat{H}_{0} + \frac{1}{\omega}\sum_n \frac{1}{n}\left[\hat{H}_{n},\hat{H}_{-n}\right],
    \label{eq:heff1st}
\end{equation}
where the operators $\hat{H}_{n}$ are the Fourier components of the time-dependent Hamiltonian \eqref{eq:tdep},
\begin{equation}
    \hat{H}_{n} = \frac{1}{T}\int_0^T \hat{H}(t) e^{-in\omega t}dt.
    \label{eq:hn}
\end{equation}
This expression is obtained within the approach discussed by Goldman and Dalibard \cite{goldman2014}, differing from the usual Magnus expansion by lacking additional commutator terms (see, e.g., Ref. \cite{bukov2015}).
We recall that since $\hat{H}_{\textrm{eff}}^{(0)}=\hat{H}_{0}$ is defined as the time average of $\hat{H}(t)$, it is characterized by the same physical processes of the undriven Hamiltonian, whose coefficients simply get renormalized, as we shall discuss in the following discussion. Instead, the structure of $\hat{H}_{\textrm{eff}}^{(1)}$ (which involves a commutator) leads -- already at first order -- to the emergence of new couplings (that is, new tunneling processes). 
This, in essence, is the reason why ``Floquet engineering'' is such a powerful tool for realizing non trivial topological models. 

Let us then consider the effective Floquet Hamiltonian defined by Eq. \eqref{eq:eff_f_h}, starting with its first-order truncation in Eq. \eqref{eq:heff1st}. 
As discussed in Ref. \cite{verdeny2015}, a bichromatic driving designed as in Eq. \eqref{opt_forc} yields \textit{isotropic} NN and next-nearest-neighbours (NNN) tunnelling rates, within the first order approximation in Eq. \eqref{eq:heff1st}. 
The leading order term $\hat{H}_{\textrm{eff}}^{(0)}=\hat{H}_{0}$ is obtained from Eq. \eqref{eq:hn}. Therefore, it is straightforward to see that the onsite energies remain unchanged (they do not depend on time), whereas the NN hopping elements get renormalized by time-averaging the expressions in Eq. \eqref{eq:tunnelings}, e.g. (we omit the label `eff' for easiness of notations), 
\begin{equation}
    t_{BA}^{\bm{0}} = \frac{j_{0}}{T}\int_0^T e^{-i\chi_{\bm{a}_3}(t)}dt.
    \label{eq:tAB}
\end{equation}
As for the first order term $\hat{H}_{\textrm{eff}}^{(1)}$, it is characterized by both a renormalization of the onsite energies 
\begin{equation}
    t_{AA}^{\bm{0}}= -t_{BB}^{\bm{0}}=\sum_{i=1}^3\beta(\bm{a}_i,-\bm{a}_i)\equiv \Delta-\delta
    \label{eq:onsite_verdeny}
\end{equation}
and, more interestingly, by the emergence of NNN hopping terms, 
\begin{equation}
    t_{AA}^{\bm{b}_i}=-t_{BB}^{\bm{b}_i} = \beta(\bm{a}_j,-\bm{a}_k),
    \label{eq:tNNNverdeny}
\end{equation}
which were vanishing in the undriven model.
In the above expressions, the indexes $i,j,k$ are cyclic integers, for example 
$t_{AA}^{\bm{b}_2} = \beta(\bm{a}_3,-\bm{a}_1)$, and the function $\beta(\bm{a}_i,\bm{a}_j)$ is defined as follows,
\begin{equation}
\beta(\bm{a}_i,\bm{a}_j) = \frac{1}{\hbar\omega}
\sum_{n=1}^{\infty}\frac{1}{n}[g_{\bm{a}_i}^{-n}g_{\bm{a}_j}^{n}
-g_{\bm{a}_j}^{-n}g_{\bm{a}_i}^{n}],  
\end{equation} 
where the functions $g_{\bm{a}_j}^{n}$ are the Fourier components of Eq. \eqref{verdeny_tun} , i.e., $g_{\bm{a}_j}^{n} = T^{-1}\int_0^Tdt\; g_{\bm{a}_j}(t)e^{-in\omega t}$ \cite{verdeny2015}.
Intuitively, one may interpret the function $\beta$ as the rate for a particle to tunnel to a site by the vector $\bm{a}_i$ and then to tunnel again through the vector $\bm{a}_j$ \cite{verdeny2015a}. As anticipated above, it can be demonstrated \cite{verdeny2015} that the NNN coefficients turn out to be isotropic, namely they are equal along the tree directions.
Therefore, the NN and NNN tunneling rates can be rewritten in a compact form, with the superscript in parenthesis denoting the order of the inverse frequency expansion the parameters are related to, as
\begin{align}
\label{eq:verdeny_t0}
t_0^{(0)} &= t_{BA}^{\bm{0}},
\\
t_1^{(1)} &= |t_1^{(1)}| e^{i\phi} = \beta(\bm{a}_i,-\bm{a}_j) \quad(i\neq j).    
\label{eq:verdeny_t1}
\end{align}
\begin{figure}[t]
\centerline{\includegraphics[width=\columnwidth]{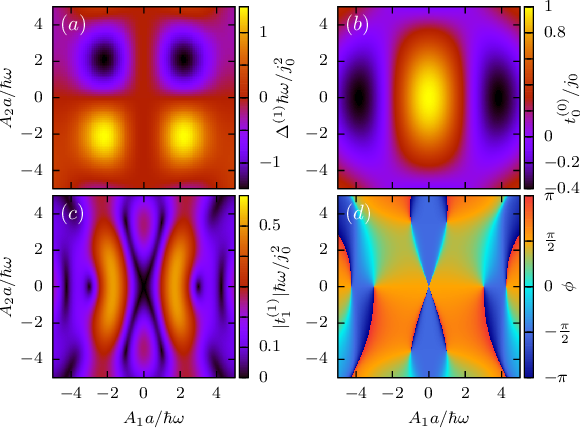}}
\caption{Tight-binding parameters of the perturbative effective Hamiltonian \eqref{eq:h3verdeny}, as a function of the amplitudes $A_1$ and $A_2$ of the driving. (a) Onsite energy $\Delta$ [Eq. \eqref{eq:onsite_verdeny}]; (b) NN tunneling amplitude $t_0^{(0)}$ [in Eq. \eqref{eq:verdeny_t0}, \eqref{eq:tAB}]; (c) and (d), modulus and phase of $t_1^{(1)}$, respectively [see Eq. \eqref{eq:verdeny_t1}]. 
Here $\hbar\omega=100\, j_0$ (fast-driving regime) and $\delta=0$ (no explicit breaking of IS).}
\label{fig:tb_vm}
\end{figure}
Then, the effective Hamiltonian $h^\textrm{eff}(\bm{k})$ reads
\begin{align}
h^\textrm{eff}_{1}(\bm{k}) &= t_0^{(0)}{\rm{Re}}[Z(\bm{k})],
\label{eq:h1verdeny}
\\
h^\textrm{eff}_{2}(\bm{k}) &= -t_0^{(0)}{\rm{Im}}[Z(\bm{k})],
\\
h^\textrm{eff}_{3}(\bm{k}) &= \Delta + 2|t_1^{(1)}|\sum_{i=1}^{3}\cos(\bm{k}\cdot\bm{b}_i+\phi).
\label{eq:h3verdeny}
\end{align}

The above results, obtained by Verdeny and Mintert in Ref. \cite{verdeny2015}, will serve as a reference for comparing the results of the numerical approach discussed in the following section.
As an example, in Fig. \ref{fig:tb_vm} we show the behavior of the three t.b. parameters,  $\Delta$, $t_0^{(0)}$, and $t_1^{(1)}$ (modulus and phase \footnote{Notice that panel (d) nicely matches Fig. 1 of Ref. \cite{verdeny2015}. An alternate version of panel (c) can be found on the preprint version of the same paper, \texttt{arXiv:1502.07350v1}, see Fig. 4 therein.}), as a function of the driving amplitudes. In particular, we consider the case in which there is no explicit IS breaking, $\delta = 0$, under fast driving conditions, with $\omega = 100j_0/\hbar$. This high-frequency value is expected to be sufficiently large to warrant the first-order truncation of the inverse frequency expansion in Eq. \eqref{eq:heff1st}.

\subsection{Non-perturbative numerical approach}
\label{sec:numerical}

In this section, we will extend the analysis beyond the lowest order in the inverse frequency expansion. To this end, we will employ the numerical approach discussed in Ref. \cite{modugno2017}.
This approach corresponds to summing up the entire inverse frequency expansion, allowing us to compute the effective Floquet Hamiltonian $\hat{H}_{F}$ without additional approximations other than those required by the numerical implementation, such as time discretization. 
We recall that $\hat{H}_{F}$ is expected to be gauge-equivalent to $\hat{H}_\textrm{eff}$ \cite{bukov2015}. 
This implies that the two Hamiltonians should share the same spectrum and topological properties. However, `local' properties, such as the tunneling coefficients, may not necessarily be equal.

To start with, we consider the effective Floquet Hamiltonian in reciprocal space, $h^{F}(\bm{k})$, that we compute by means of a suitable time discretization, as follows 
\begin{equation}
h^{F}(\bm{k})=\frac{i}{T}\ln\left(\prod_{i=0}^{N-1}e^{-i\Delta_{t}h(\bm{k},t_{i})}\right)_{\cal T},
\label{eq:hkf}
\end{equation}
with $\Delta_{t}\equiv T/N$. The details of numerical scheme are summarized in Appendix \ref{appendix2}.
Then, the tunneling coefficients can be easily obtained as
\begin{equation}
t_{\nu\nu'}^{\bm{\ell}}=\frac{1}{\Omega_{\cal B}}\int_{\cal B} 
h_{\nu\nu'}^F(\bm{k})e^{-i\bm{k}\cdot\bm{R}_{\bm{\ell}}}d\bm{k},
\label{eq:nptunnelings}
\end{equation}
see Appendix \ref{appendix1}.
Consistently with the notations in Sec. \ref{sec:perturbative}, we indicate the non-perturbative ($np$) effective onsite energies, as well as the NN and NNN tunnelling rates obtained through this method as $\Delta_{\nu}^{np}$, $t_{0i}^{np}$, and $t_{1{\nu},i}^{np} = |t_{1{\nu},i}^{np}|e^{i\phi_{{\nu}i}^{np}}$ ($\nu=A,B$), respectively. The index $i$, introduced to account for any possible non-isotropy, matches that of the vectors $\bm{a}_i$ in case of NN tunneling rates ($i=1,2,3$), and of $\bm{b}_i$ for NNN hoppings (with $i=4,5,6$ for $-\bm{b}_i$), see Fig. \ref{fig:honeycomb}. 

The behavior of the various tight-binding parameters of the model as a function of the amplitudes $A_1$ and $A_2$ of the driving is discussed below. We focus on the same scenario as considered in Section \ref{sec:perturbative}, specifically addressing the case where there is no explicit IS breaking, $\delta = 0$, under fast driving conditions, with $\omega = 100j_0/\hbar$.

\begin{figure}[t]
\centerline{\includegraphics[width=\columnwidth]{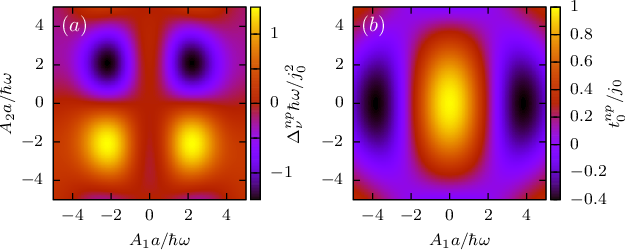}}
\caption{(a) Onsite energy $\Delta^{np}$ and (b) NN tunneling amplitude $t_0^{np}$, obtained from the non-perturbative numerical approach (see text), as a function of the amplitudes $A_1$ and $A_2$ of the driving.
Here $\hbar\omega=100\, j_0$ (fast-driving regime) and $\delta=0$ (no explicit breaking of IS).}
\label{fig:onsite-t0}
\end{figure}
We begin by illustrating the behavior of the onsite energies and of the NN tunneling coefficients.
Regarding the former, we observe that the two lattice sites exhibit opposite onsite energies, as obtained in the model of Ref. \cite{verdeny2015}, see Eq. \eqref{eq:onsite_verdeny}.
Therefore, moving forward, we can simplify our notation by setting $\Delta^{np}\equiv\Delta_{A}^{np}=-\Delta_{B}^{np}$. This quantity is shown in Fig. \ref{fig:onsite-t0}a, which nicely matches with Fig. \ref{fig:tb_vm}a. These figures clearly reveal the dynamical breaking of the degeneracy between the two lattice sites caused the driving.

Similarly, we find that the NN tunnelling rate $t_{0i}^{np}$ is isotropic, as in the non-stroboscopic model, so that the index $i$ can be safely omitted. Its behavior, shown in Fig. \ref{fig:onsite-t0}b, perfectly reproduces that displayed in Fig. \ref{fig:tb_vm}b, proving that also the emergent NN tunneling is model-independent.
This was expected because in the fast-driving regime the value of the NN tunneling coefficient is determined by the zeroth order effective Hamiltonian, where both the stroboscopic and non-stroboscopic have the same expression as in Eq. \eqref{eq:hn} (with $n=0$), and therefore they coincide. This provides a robust characterization of the system's topological properties, as from the isotropy of the NN hoppings it follows that the position of the Dirac points is fixed by the space group of the model, see Appendix \ref{appendix3}. This has to be a gauge invariant property, so that it cannot depend on the model, as confirmed by the numerical results.

\begin{figure}[t]
\centerline{\includegraphics[width=\columnwidth]{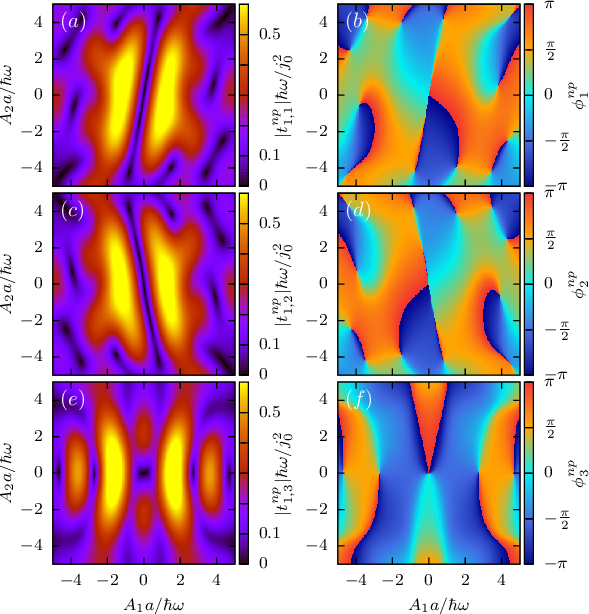}}
\caption{Behavior of the emergent NNN tunneling rates $t_{1i}^{np}$ obtained from the non-perturbative numerical approach (see text), plotted as a function of the amplitudes $A_1$ and $A_2$ of the driving. The modulus and phase are presented in the left and right columns, respectively, for $i = 1$ ($a, b$), $i = 2$ ($c, d$), and $i = 3$ ($e, f$).
Here $\hbar\omega=100\, j_0$ (fast-driving regime) and $\delta=0$ (no explicit breaking of IS).}
\label{fig:t1}
\end{figure}

Then, in Fig. \ref{fig:t1} we show the modulus of the emergent NNN tunnelling rates, $|t_{1\nu,i}^{np}|$, and their complex phases $\phi_{{\nu}i}^{np}$, in the left and right columns respectively.
Consistently with the property in Eq. \eqref{eq:tNNNverdeny} found from the perturbative approach \footnote{As discussed in Ref. \cite{verdeny2015,verdeny2015a}, this is a fundamental symmetry of the model, independent of the specific design of the driving, that can be traced back to the vanishing of the NNN tunnelings of the undriven model. It is worth reminding that instead, in the Haldane model, the NNN tunneling rates for sites $A$ and $B$ comes in conjugate pairs, $t_{1B,1}^{np}=t_{1A,1}^{np^*}$.}, the numerical results obey the relation $t_{1B,1}^{np}=-t_{1A,1}^{np}$. Therefore, in the figure and in the following discussion the $\nu$ index is omitted, for the sake of notation simplicity.

On the other hand, we find that the NNN tunneling rates are model-dependent, both in modulus and phase, see Figs. \ref{fig:tb_vm}c,d for comparison. 
In particular, in addition to being inherently dependent on the initial time $t_{0}$, the results obtained by means of the stroboscopic Floquet approach turn out to be anisotropic, with mirrored NNN neighbors constituting complex conjugate pairs, namely $t_{1,i+3}^{np} = t_{1,i}^{np*}$. 
Interestingly, we have confirmed that isotropy cannot be restored, even by considering a different initial time $t_{0}$, akin to the findings in Ref. \cite{modugno2017}. 
However, as we shall explicitly compute in the following section, these `local' features of the specific effective Hamiltonian representation are not relevant as far as the topological properties of the system are concerned. In addition, despite the different values that the NNN tunneling rates may take within each approach, it is worth noticing that they follow the predicted scaling behavior with respect to $j_0^2/\omega$, as it is evident from the color scale of the figure. This provides further validation of the theoretical framework.

It is worth noting that, upon numerically computing the Floquet Hamiltonian from Eq. \eqref{eq:hkf}, one can easily extract all tunneling coefficients at any order within the tight-binding expansion using Eq. \eqref{eq:nptunnelings}. 
We have verified that the tunneling coefficients beyond the NNN order are indeed negligible. Therefore, in the current fast-driving scenario, the tight-binding expansion can be safely truncated at that order. Then, by using Eq. \eqref{eq:h_momentum_space}, the Floquet Hamiltonian can be analytically expressed as
\begin{align}
h_{1}^F(\bm{k}) &= t_{0}^{np}{\rm{Re}}[Z(\bm{k})],
\\
h_{2}^F(\bm{k}) &= -t_{0}^{np}{\rm{Im}}[Z(\bm{k})],
\\
h_{3}^F(\bm{k}) &= \Delta^{np} + 2\sum_{i=1}^{3}|t_{1,i}^{np}|\cos(\bm{k}\cdot\bm{b}_{i}+\phi_{i}^{np}),
\label{eq:h3np}
\end{align}
again with $h_{0}^F(\bm{k})\equiv0$. Here $\Delta^{np}$ includes also the explicit energy offset $\delta$ between sites $A$ and $B$.

\section{Topological phase diagram}
\label{sec:topology}

We will now turn to analyze the topological properties of the system, considering both the case analyzed so far, with $\delta=0$, and the one in which IS is \textit{explicitly broken}, $\delta \neq 0$.
In particular, we will draw the topological phase diagram as a function of the driving amplitudes $A_1$ and $A_2$ (for $\delta = 0$) or as a function of  $A_1$ and $\delta$, by keeping $A_2$ fixed (for $\delta \neq 0$). 

\begin{figure}[t!]
\centerline{\includegraphics[width=\columnwidth]{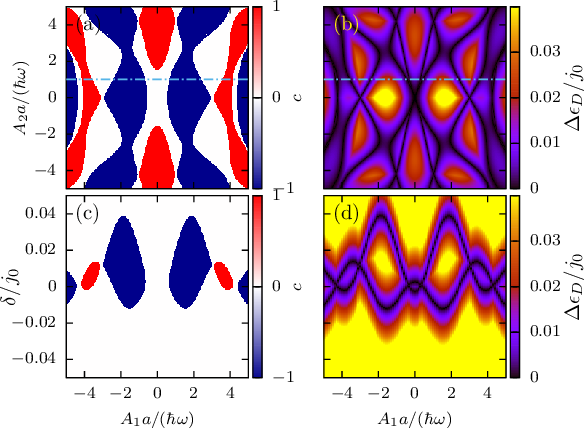}}
\caption{Topological phase diagram
in the fast-driving regime  $\hbar\omega=100\, j_0$. Panels (a,c) display the Chern number $c$ obtained from the numerical non-perturbative approach:  
(a) Inversion-symmetric case, $\delta = 0$, as a function of the driving amplitudes $A_1$ and $A_2$; 
(b) Explicit breaking of IS, as a function of $A_1$ and of the bare onsite energy gap $\delta$,for $A_{2}a/(\hbar\omega)=1$ [as indicated to the dot-dashed line in panel (a)].
Panels (c,d) show the corresponding behavior for the minimal energy gap at the Dirac points, $\Delta\epsilon_{D}$ (see text).}
\label{fig:topological_pd}
\end{figure}

The central quantity needed to discuss the topological properties of the system is the Chern number $c$. For a given band $n$, the corresponding Chern number can be written as \cite{ibanez2015,modugno2017}
\begin{equation}
    c_n = \frac{1}{2\pi}\int_{\cal B}\Omega_n(\bm{k})\, d\bm{k},
    \label{eq:chern}
\end{equation}
where $\Omega_n(\bm{k})$ is the corresponding Berry curvature
\begin{equation}
    \Omega_n(\bm{k}) = 2\text{Im}\sum_{m\neq n}\frac{\braket{n|h_x(\bm{k})|m}\braket{m|h_y(\bm{k})|n}}{[\epsilon_m(\bm{k})-\epsilon_n(\bm{k})]^2},
    \label{eq:berrycurv}
\end{equation}
with $\epsilon_n(\bm{k})$, $\ket{n(\bm{k})}$ being the eigenvalues and eigenstates of the Hamiltonian $h(\bm{k})$, while $h_\alpha(\bm{k})$ is a shorthand notation for $\partial h(\bm{k})/\partial k_\alpha$ ($\alpha=x,y$). In the present context of a two-band problem where solely the lowest band is supposed to be occupied, only the knowledge of $c\equiv c_1$ is required. Therefore, in Eq. \eqref{eq:berrycurv}, one can set $n = 1$ and $m = 2$. In the numerical calculations, we find convenient to tile the reciprocal space with rhomboidal cells containing the two inequivalent Dirac points, akin to the coordinate space tiling shown in Fig. \ref{fig:honeycomb}, see Appendix \ref{appendix3}. This choice proves particularly effective for an accurate computation of the Chern number since the main contribution to the integral comes from the proximity of the Dirac points, where the Berry curvature $\Omega_n(\bm{k})$ is peaked (see also Ref. \cite{modugno2017}). In addition, we have also verified that both quasienergies $\epsilon_n(\bm{k})$ ($n=1,2)$ never wrap around the edges of the \textit{quasienergy ﬁrst Brillouin zone} \cite{holthaus2015,pieplow2018}, so that the denominator of Eq. \eqref{eq:berrycurv} can be computed with no ambiguity. 

The topological phase diagram obtained from Eq. \eqref{eq:chern} is shown in the left panels of Fig. \ref{fig:topological_pd}, for the same fast-driving regime $\hbar\omega=100\, j_0$ considered in the previous section. 
It is characterized by a very rich structure, in which topologically nontrivial phases with $c=\pm1$ alternate as a consequence of the interplay between IS and TRS dynamical breaking. We recall that the current driving protocol does not exhibit IS, so that IS gets broken even in the case in which the onsite energies of undriven model are not degenerate ($\delta=0$), see Fig. \ref{fig:topological_pd}a. This would not be possible in case of a monochromatic driving, like the one considered in Refs. \cite{jotzu2014, modugno2017}, where an explicit breaking of IS is necessary in order to reproduce the full Haldane topological phase diagram.

Along with the phase diagram, in the right panels of Fig. \ref{fig:topological_pd} we show the minimal energy gap at the Dirac points, defined as $\Delta\epsilon_{D}\equiv\min(\epsilon_2(\bm{k})-\epsilon_1(\bm{k}))|_{\bm{k}=\bm{k}_{D}^\pm}$. This quantity is specially relevant, considering that the boundaries between the different topological phases correspond to the vanishing of the gap at one of the two inequivalent Dirac points. In particular, the thick black lines in Fig. \ref{fig:topological_pd}b,d correspond to the regions where $\Delta\epsilon_{D}=0$. It is worth noticing that, though they exhibit a finite width due to numerical precision, they nicely match the phase boundaries in Figs. \ref{fig:topological_pd}a,c. This further confirms the robustness of the present results.

It is worth emphasising that here we have opted to construct the topological phase diagram based on the physical parameters characterizing the driving ($A_1$, $A_2$) and the original Hamiltonian ($\delta$), rather than relying on the conventional approach found in the literature, which employs tight-binding parameters such as the tunneling amplitude $|t_1|$ and its phase $\phi$ \cite{verdeny2015,haldane1988} (see also the discussion in Ref. \cite{ibanez2015}). This choice is motivated by two key considerations. Firstly, the amplitudes ($A_1,\,A_2$) are the parameters that define the driving protocol, and as such they are directly accessible and tunable in any experimental implementation, making them the natural choice from the experimental point of view. Secondly, as we have seen in the previous section, the values of the tunneling parameters are inherently model-dependent, introducing complexity and potential ambiguity into the interpretation of results. 

Nevertheless, it is important to remark that the two approaches for constructing the topological phase diagram are equivalent in terms of physical content. Indeed, we have verified that the topological phases predicted by the NNN truncated effective approach, as presented in Fig. 2 of Ref. \cite{verdeny2015}, can be mapped onto the phase diagram in Fig. \ref{fig:topological_pd}a. This can be conveniently demonstrated by determining the values of $\Delta(A_1,A_2)$ from Eq. \eqref{eq:onsite_verdeny}, along with $|t_1^{(1)}|(A_1,A_2)$ and $\phi(A_1,A_2)$ from Eq. \eqref{eq:verdeny_t1}, and then associating the corresponding Chern numbers shown in Ref. \cite{verdeny2015} (alternatively, the Chern number can be computed as discussed in  Ref. \cite{sticlet2012}). As anticipated in the introduction, this result provides a direct confirmation of the fact that the topological properties of the driven model do not depend on the specific approach employed to compute them, either the \textit{effective} Hamiltonian $\hat{H}_{\textrm{eff}}$ or the \textit{stroboscopic} Floquet Hamiltonian $\hat{H}_{F}[t_0]$.

\section{Conclusions}
\label{sec:conclusions}

By means of a direct non-perturbative numerical approach, we have computed the stroboscopic Floquet Hamiltonian and the topological phase diagram of a system of non-interacting ultracold atoms in a shaken honeycomb lattice, subjected to the optimal driving of Ref. \cite{verdeny2015}. 
Such a driving induces dynamical breaking of inversion and time-reversal symmetries, thus providing an effective way to reproduce the topology of the Haldane model, to some extent. 
The numerical results have been compared with the analytical prediction of the non-stroboscopic effective approach of Ref. \cite{verdeny2015}, valid in the fast-driving regime, with the objective to provide a general discussion of the different perspectives for computing the effective Floquet Hamiltonian of periodically driven systems.

We have found that the topological properties are indeed gauge invariant, as expected. However, some of the tunneling parameters are model-dependent. Specifically, we confirm that the emergent NN tunneling rates are isotropic \cite{verdeny2015}, meaning that the position of the Dirac points is fixed solely by the lattice geometry and is, therefore, model-independent. On the other hand, the NNN tunneling rates turn out to be anisotropic within the stroboscopic Hamiltonian approach, in contrast to the non-stroboscopic case \cite{verdeny2015}.
Therefore, we argue that the topological phase diagram is most effectively represented in terms of the experimentally controllable physical parameters defining the driven Hamiltonian, rather than the model-dependent emergent tunneling parameters—such as the next-nearest-neighbor tunneling amplitude $|t_1|$ and its phase $\phi$ \cite{verdeny2015, haldane1988, ibanez2015}.

Finally, we emphasize that the Floquet Hamiltonian approach represents a very valuable complementary approach to the gauge-invariant representation provided by the effective Hamiltonian of Ref. \cite{goldman2014}. 
In fact, in the case of a non-interacting system, it enables a direct non-perturbative calculation using simple numerical methods.
In this respect, the fact that it is intrinsically dependent on the initial time $t_0$ is only a minor shortcoming, as all physical quantities are gauge invariant and can be computed without ambiguities.

\begin{acknowledgments}
We acknowledge support from Grants No. PID2021-126273NB-I00 and No. PID2021-129035NB-I00 funded by MCIN/AEI/10.13039/501100011033 and by ``ERDF A way of making Europe'', from the Basque Government through Grant No.  IT1470-22, and from from the European Union’s Horizon 2020 research and innovation programme under the European Research Council (ERC) grant agreement No 946629.
\end{acknowledgments}

\appendix

\section{Momentum representation of the Hamiltonian and tunneling coefficients}
\label{appendix1}

Let us consider the general expression of a tight-binding single-particle Hamiltonian, as in Eq. \eqref{eq:und_hamil},
\begin{equation}
\hat{H} = \sum_{\nu,\nu'}\sum_{\bm{j},\bm{j}'}{\hat{c}}^{\dagger}_{\bm{j}\nu}{\hat{c}}_{\bm{j}'\nu'}t_{\bm{j}-\bm{j}'}^{\nu\nu'}.
\label{eq:hamiltonian}
\end{equation}
The latter expression can be mapped in reciprocal space, as follows \cite{modugno2016}. 
We start by defining 
\begin{equation}
\hat{c}_{\bm{j}\nu}=\frac{1}{\sqrt{\Omega_{\cal B}}}\int_{\cal B} \hat{d}_{\nu\bm{k}}e^{-i\bm{k}\cdot\bm{R}_{\bm{j}}}d\bm{k},
\end{equation}
where $\cal B$ refers to the first Brillouin zone, $\Omega_{\cal B}$ its $d-$volume, and $d$ the dimensionality of the system.  
Then, the Hamiltonian \eqref{eq:hamiltonian} can be written as 
\begin{equation}
\hat{H}=\sum_{\nu\nu'}\int_{\cal B} h_{\nu\nu'}(\bm{k})
\hat{d}_{\nu\bm{k}}^{\dagger}\hat{d}_{\nu'\bm{k}}d\bm{k},
\end{equation}
with 
\begin{equation}
h_{\nu\nu'}(\bm{k})=\sum_{\bm{\ell}}t_{\nu\nu'}^{\bm{\ell}}e^{i{\bm{k}}\cdot{\bm{R}}_{\bm{\ell}}},
\label{eq:h_momentum_space}
\end{equation}
where the lattice index $\bm{\ell}$ identifies the position of each cell with respect to the central one.
By inverting the latter expression, it is straightforward to show that 
the tunneling coefficients can be expressed as
\begin{equation}
t_{\nu\nu'}^{\bm{\ell}}=\frac{1}{\Omega_{\cal B}}\int_{\cal B} 
h_{\nu\nu'}(\bm{k})e^{-i\bm{k}\cdot\bm{R}_{\bm{\ell}}}d\bm{k}.
\end{equation}

\section{How to compute numerically the effective Hamiltonian}
\label{appendix2}

The effective Hamiltonian in Eq. \eqref{eq:hkf} can be computed as follows. 
We define (we omit the dependence depends on $\bm{k}$ for easiness of notations)
\begin{equation}
u\cdot\tilde{\sigma}\equiv U(t_{i}+\Delta_{t},t_{i})=e^{-i\Delta_{t}h(\bm{k},t_{i})}\equiv e^{-i\eta\cdot\tilde{\sigma}},
\end{equation}
where we have used the representation of $2\times2$ matrices on the basis of Pauli matrices, 
$\eta=(\eta_{0},\eta_{1},\eta_{2},\eta_{3})\equiv(\eta_{0},\bm{\eta})$, and similarly for $u$. The quantities
$u$ and $\eta$ are connected by the following relations 
\begin{align}
u&=e^{-i\eta_{0}}\left(\cos|\bm{\eta}|, -i\frac{\bm{\eta}}{|\bm{\eta}|}\sin|\bm{\eta}|\right),
\\
\eta&=\left(\eta_{0},ie^{i\eta_{0}}\frac{|\bm{\eta}|}{\sin|\bm{\eta}|}\bm{u}\right),
\end{align}
with $\eta_{0}=-\arg(u_{0})$, $|\bm{\eta}|=\cos^{-1}|u_{0}|$.
The total evolution operator over one period is computed as
\begin{equation}
U(t_{0}+T,t_{0})=u_{N-1}\cdot u_{N-2}\cdot\dots\cdot u_{1}\cdot u_{0}\equiv u
\end{equation}
and the effective Hamiltonian as $h_{\textrm{eff}}={\eta}/{T}$.

\section{Dirac points}
\label{appendix3}

\begin{figure}[b]
	\centerline{\includegraphics[width=\columnwidth]{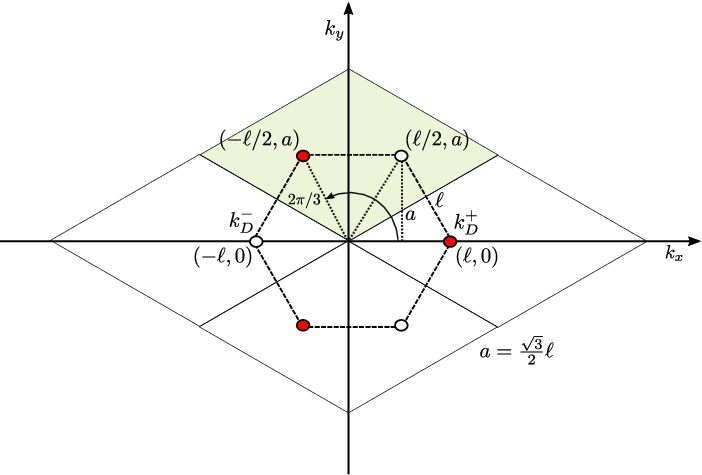}}
	\caption{Sketch of the geometry in reciprocal space and of the configuration of the Dirac points. 
    The diamond-shaped area shaded in light gray represents a suitable choice for the first Brillouin zone.}
	\label{fig:diracpoints}
\end{figure}

A spectrum of the form as in Eq. \eqref{eq:spectrum}, $\varepsilon_\pm =\pm\sqrt{\sum_{i}h_{i}^{2}(\bm{k})}$, may be characterized by the so-called Dirac points $\bm{k}=\bm{k}_{D}$, at which the two bands become degenerate, and their local dispersion is linear (corresponding to relativistic particles with vanishing mass) \cite{lee2009}. Their positions are fixed by the condition
\begin{equation}
h_{1}^{2}(\bm{k}_{D})+h_{2}^{2}(\bm{k}_{D}) = 0 \Rightarrow |Z(\bm{k}_{D})|=0,
\label{eq:kd}
\end{equation}
which depend only on the geometry of the system. For this specific lattice geometry, there are two inequivalent Dirac points, which are the dual of the two lattice sites. 
The vanishing of the gap takes place instead when also $h_{3}(\bm{k}_{D})=0$ at any of the two Dirac points. This may occur for some specific values of the  NNN tunneling rate and of the onsite energy difference, see Eqs. \eqref{eq:h3verdeny} and \eqref{eq:h3np} [not for the undriven model, see Eq. \eqref{eq:h3undriven}]. 
The solutions of Eq. \eqref{eq:kd},
\begin{equation}
Z(\bm{k}_{D})=1 + e^{i\bm{k}_{D}\cdot\bm{b}_{1}} + e^{-i\bm{k}_{D}\cdot\bm{b}_{2}}=0,
\end{equation}
correspond to $\bm{k}_{D}\cdot\bm{b}_{1}=\pm{2\pi}/{3}=\bm{k}_{D}\cdot\bm{b}_{2}$, modulo integer multiples of $\pm2\pi$. Then, it is straightforward to get
\begin{equation}
\bm{k}_{D}^{\pm}=\frac{4\pi}{3\sqrt{3}a}\left(\pm1,0\right).
\end{equation}
As anticipated, other solutions are also possible. Some of them are shown in Fig. \ref{fig:diracpoints}, along with the conventional tiling in Brillouin zones.

%

\end{document}